\documentclass[12pt]{article}
\usepackage{amsmath}
\usepackage{amssymb}
\usepackage{amstext}
\usepackage{amsfonts}
\usepackage{graphicx,epsfig}
\usepackage{slashbox}
\usepackage{indentfirst}
\newcommand{\calB}{{\cal B}}
\newcommand{\calL}{{\cal L}}

\newcommand{\calP}{{\cal P}}
\newcommand{\calM}{{\cal M}}

\newcommand{\calA}{{\cal A}}

\newcommand{\kslash}{/\!\!\!k}
\newcommand{\pslash}{/\!\!\!p}
\newcommand{\qslash}{/\!\!\!q}

\begin{document}

\baselineskip 20pt

\title{Study on the effects of the light CP-odd Higgs via the
leptonic decays of pseudoscalar mesons }

\author{Liang Tang$^a$, Hong-Wei Ke$^b$ and Xue-Qian Li$^a$\\[0.5cm]
{\small $a)$ School of Physics, Nankai University, 300071, Tianjin,
China}\\
\small $b)$ School of Science, Tianjin University, 300072, Tianjin,
China}
\date{}
\maketitle

\noindent {\bf Abstract}

To explain the anomalously large decay rate of $\Sigma^+\to
p+\mu^+\mu^-$, it was proposed that a new mechanism where a light
CP-odd pseudoscalar boson of $m_{A_1^0}=214.3$ MeV makes a crucial
contribution. Later, some authors have studied the transition
$\pi^0\to e^+e^-$ and $\Upsilon\to \gamma A_1^0$ in terms of the
same mechanism and their result indicates that with the suggested
mass one cannot fit the data. This discrepancy might be caused by
experimental error of $\Sigma^+\to p+\mu^+\mu^-$ because there were
only a few events. Whether the mechanism is a reasonable one
motivates us to investigate the transitions $\pi^0\to e^+e^-;\;\eta
(\eta^\prime)\to \mu^+\mu^-;\; \eta_c\to
\mu^+\mu^-;\;\eta_b\to\tau^+\tau^-$ within the same framework. It is
noted that for $\pi^0\to e^+e^-$, the standard model (SM) prediction
is smaller than the data, whereas the experimental central value of
$\eta \to \mu^+\mu^-$ is also above the SM prediction. It means that
there should be extra contributions from other mechanisms and the
contribution of $A_1^0$ may be a possible one. Theoretically
calculating the branching ratios of the concerned modes, we would
check if we can obtain a universal mass for $A_1^0$ which reconcile
the theoretical predictions and data for all the modes.
Unfortunately, we find that it is impossible to have such a mass
with the same coupling $|g_\ell|$. Therefore we conclude that the
phenomenology does not favor such a light $A_1^0$, even though a
small window is still open.

\noindent{PACS numbers:
12.60.Jv, 
13.20.Gd, 
13.35.-r, 
14.80.Da. 
}

\section{Introduction}

Searching for new physics beyond the standard model (SM) is the goal
of not only the very high energy experiments such as at LHC and even
the future ILC, but also the machines of lower energies where new
physics signals may be revealed at rare processes. The HyperCP
collaboration observed an anomalously large decay rate of $\Sigma\to
p+\mu^+\mu^-$ which is higher than the prediction of SM by several
standard deviations\cite{Park:2005eka}. The discrepancy may be
attributed to contributions from new physics. A very possible
mechanism is that the extra contribution is due to a light CP-odd
pseudoscalar. Indeed, many models suggests its existence. Among all
the models the supersymmetric model where there are five Higgs
bosons remain after the symmetry breaking is the most favorable one.
Namely there are two CP-even scalars $H^0$ and $h^0$, a CP-odd $A^0$
and two charged Higgs bosons $H^{\pm}$\cite{Rosiek:1989rs}.

The search for SM Higgs boson has already spanned for almost half
century and covered a rather large regions. Recently, at LHC an
excess at 126 GeV is observed and it could be the signal of Higgs,
even though firm identification still needs
time\cite{114,ATLAS,CMS}. So far, as well known, the SM Higgs must
be heavy, but a Higgs demanded by new physics beyond the SM might be
light. Phenomenological search for beyond SM Higgs would be an
interesting job of theorists and experimentalists.  For example, Kao
{\it et al} investigate the FCNC process $t\to c\phi$ with the two
Higgs Doublet Model  at  LHC \cite{Kao:2011aa} where a heavy scalar
or pseudoscalar Higgs $\phi$ of about 130 GeV contributes. On other
aspect, it does not exclude the possibility that a light CP-odd
pseudoscalar boson might exist, but it definitely is not the SM
Higgs. To explain the large decay rate $\Sigma\to p+\mu^+\mu^-$
which should be very small if only the SM applies, He, Tandean and
Valencia \cite{He:2006fr} suggested that a light CP-odd $A_1^0$ of
mass 214.3 MeV may result in the observed data. Later Chang and Yang
applies the same mechanism to evaluate the branching ratios of
$\pi^0\to e^+e^-$ while considering a constraint from
$\calB(\Upsilon\to \gamma A_1^0)$ \cite{Chang:2008np}. They noticed
that the SM prediction is lower than the experimental
data\cite{Abouzaid:2006kk}, therefore there should be some extra
contributions from the mechanisms which have not been considered yet
or are due to new physics beyond SM. Chang and Yang calculated the
contribution of the light $A_1^0$, but combining the constraints
from the anomalous magnetic moment of muon and $\calB(\Upsilon\to
\gamma A_1^0)$ on $A_1^0$, the authors concluded that rigorous
constraints on the mass of $A_1^0$ and the concerned parameter
$|g_\ell|$ enforced by $\calB(\pi^0\to e^+e^-)$ and
$\calB(\Upsilon\to \gamma A_1^0)$ rule out the mass of 214.3 MeV.
Furthermore by fitting data, if the mechanism does make a
substantial contribution, one should have $m_{A_1^0}\sim m_{\pi}$
and $|g_\ell|=0.10\pm 0.08$.

Considering that the HyperCP collaboration only recorded a few
events for $\Sigma\to p+\mu^+\mu^-$, thus relatively large
experimental uncertainties could be expected, it is natural to ask
if a mass range of $A_1^0$ at vicinity of $m_{\pi}$ can remarkably
enhance the rate of $\Sigma\to p+\mu^+\mu^-$? Moreover, since the
new contribution to $\pi^0\to e^+e^-$ is realized via mediating a
Higgs-like boson in the s-channel, its coupling to the lepton is
proportional to its mass, so that the contribution is suppressed by
the electron mass. And due to the phase space restriction, $\pi^0$
cannot decay into heavier muons.

In fact, we notice that for similar decay modes, the SM predictions
on $\eta\to \mu^+\mu^-$ is below the experimental central value,
\footnote{The experimental error for $\calB(\eta\to \mu^+\mu^-)$ is
large, so that making a definite conclusion needs more precise
measurement which will be coming soon. }  and there are no data
available yet for the modes $\eta^\prime\to \mu^+\mu^- \; \eta_c\to
\mu^+\mu^-, \; \eta_b\to \tau^+\tau^-$ etc., and we will show below
that they are important for determining if the scenario of $A_1^0$
works. The observation may hint that there could exist an unknown
mechanism(s) which can make up the gap between the SM prediction and
data. Existence of a light $A_1^0$ definitely is a reasonable
candidate. Thus in this work, we are going to carry out a wider
study on the the modes in terms of the theory which involves a light
CP-odd boson $A_1^0$ originating from the NMSSM theory\cite{NMSSM}
and the concerned couplings with fermions is:
\begin{eqnarray}
\calL_{A_1^0qq}&=&-\left(\sum_{\text{u-type}}l_um_u\bar{u}\gamma_5
u+\sum_{\text{d-type}}l_dm_d\bar{d}\gamma_5
d\right)\frac{iA_1^0}{v}\;,\label{NMSSM-Lagrangian-1}\\
\calL_{A1^0\ell\ell}&=&\frac{ig_\ell
m_\ell}{v}\bar{\ell}\gamma_5\ell A_1^0\;,\label{NMSSM-Lagrangian-2}
\end{eqnarray}
where, $l_d=-g_\ell=v\delta_{-}/(\sqrt{2}x)$ and
$l_u=l_d/\tan^2\beta$\cite{He:2006fr}.

Our strategy is whether we can find a mass range as well as the
parameter $|g_\ell|$ (see the text for detail), which can tolerate
all the observed modes, namely a universal $A_1^0$ mass can make the
gaps between the SM predictions and the data.

The paper is arranged as follows. In section 2, we present the
necessary theoretical derivations. In section 3, our numerical
results are shown in relevant tables and figures. We reserve the
last section for our discussion and conclusion.

\section{Formalism}
To serve our aim of this work, we concentrate ourselves on the
application of the light CP-odd pseudoscalar boson $A_1^0$ in the
NMSSM. Since the leptonic decays of the pseudoscalar mesons
$\pi^0,\;\eta,\;\eta^\prime,\;\eta_c,\;\eta_b$ are less contaminated
by the non-perturbative QCD effects, they are ideal for studying the
new mechanism. As aforementioned, unlike $\pi^0\to e^+e^-$, the
contributions of $A_1^0$ to the decay modes
$\eta(\eta^\prime)\to\mu^+\mu^-$, $\eta_c\to\mu^+\mu^-$ and
$\eta_b\to\tau^+\tau^-$ do not severely suffer from the mass
suppression. Then, pre-assuming the new mechanism, by fitting data
we would check if we can obtain a universal mass for $A_1^0$ which
reconciles all the modes.
\subsection{For leptonic decays of light pseudoscalar mesons}
\begin{figure}
\centering
\includegraphics[width=6cm]{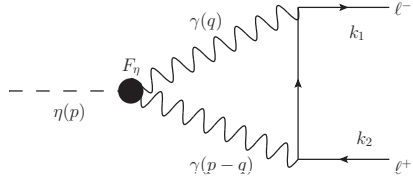}
\caption{The QED contribution to the $\eta\to\ell^+\ell^-$
decay}\label{oneloop-triangle}
\end{figure}

In the SM sector, the dominant contribution to
$\eta\rightarrow\mu^+\mu^-$ comes from the QED anomaly and the
Feynman diagram is shown in Fig.(\ref{oneloop-triangle}).
For completeness, we re-derive the formulae given in
Ref.\cite{Chang:2008np,Silagadze:2006rt} and show them in this text.
The total contribution of the triangle-diagram is written as:
\begin{eqnarray}
\calM_{\gamma\gamma}=ie^2\int\frac{d^Dq}{(2\pi)^D}\frac{L^{\mu\nu}H_{\mu\nu}}{(q^2+i\varepsilon)((q-p)^2+i\varepsilon)
((q-k_1)^2-m_\ell+i\varepsilon)}\;,
\end{eqnarray}
where
\begin{eqnarray}
L^{\mu\nu}&=&\bar{u}(k_1,s)\gamma^\mu(\kslash_1-\qslash+m_\ell)\gamma^\nu
v(k_2,s^\prime)\;,\\
H_{\mu\nu}&=&ie^2\epsilon_{\mu\nu\alpha\beta}q^\alpha(p-q)^\beta
f_{\gamma\gamma}F_{\eta\gamma\gamma}(q^2,(p-q)^2)\;.
\end{eqnarray}
Explicitly, $H_{\mu\nu}$ is the effective $\eta\gamma\gamma$ vertex
where the Lorentz structure includes a form factor related to the
loop integral. The form factor, as usual, can be decomposed into a
numerical coupling constant $f_{\gamma\gamma}=1/(4\pi^2 f_{\eta})$
times a function $F_{\eta\gamma\gamma}(q^2,(p-q)^2)$.

For the lepton part $L^{\mu\nu}$, we employ the projection operator
technique \cite{Silagadze:2006rt}:
\begin{eqnarray}
\calP(p-k_1,k_1)&=&\frac{1}{\sqrt{2}}[v(p-k_1,+)\otimes\bar{u}(k_1,-)+v(p-k_1,-)\otimes\bar{u}(k_1,+)]\nonumber\\
&=&\frac{1}{2\sqrt{2p^2}}[-2m_\ell
p_\mu\gamma^\mu\gamma^5+\frac{1}{2}\epsilon_{\mu\nu\sigma\tau}\left(k_1^\sigma(p-k_1)^\tau-(p-k_1)^\sigma
p^\tau\right)\sigma^{\mu\nu}\nonumber\\
&+&p^2\gamma^5]\;.
\end{eqnarray}

Then we have the amplitude\footnote{Here, we have added the missing
minus according the corrections to M. E. Peskin's QFT book:
\text{http://www.slac.stanford.edu/~mpeskin/QFT.html\#errors}. } for
Fig.(\ref{oneloop-triangle}):
\begin{eqnarray}
\calM_{\gamma\gamma}(\eta\to\ell^+\ell^-)=-2\sqrt{2}\alpha_{em}^2m_{\ell}m_\eta
f_{\gamma\gamma}\calA^{\ell}(m_\eta^2)\;.
\end{eqnarray}
where $\calA^{\ell}(p^2)$ is the reduced amplitude:
\begin{eqnarray}\label{reduced-amplitude}
\calA^{\ell}(p^2)=\frac{2i}{p^2}\int\frac{d^4q}{\pi^2}\frac{q^2p^2-(q\cdot
p)^2}{(q^2+i\varepsilon)((q-p)^2+i\varepsilon)((q-k_1)^2-m_\ell^2+i\varepsilon)}F_{\eta\gamma\gamma}(q^2,(p-q)^2)\;.
\end{eqnarray}
This is the same as the result given in Ref.\cite{Silagadze:2006rt},
and in the derivation, we also utilize the same form factor
$F_{\eta\gamma\gamma}(q^2,(p-q)^2)$ therein.

\begin{figure}
\centering
\includegraphics[width=6cm]{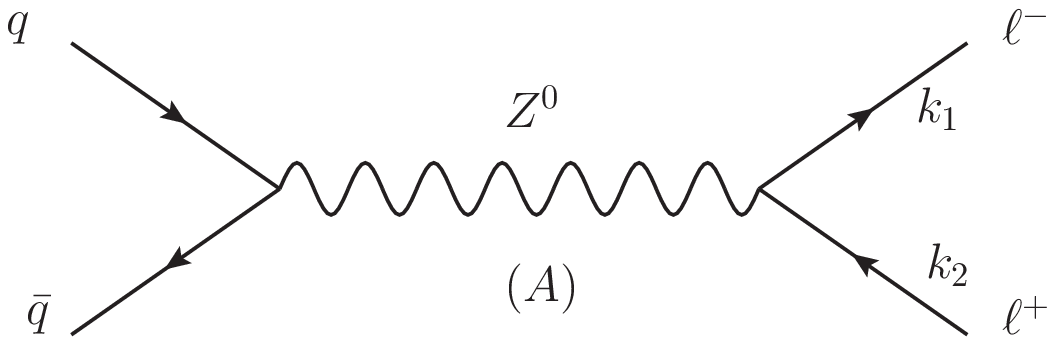}
\includegraphics[width=6cm]{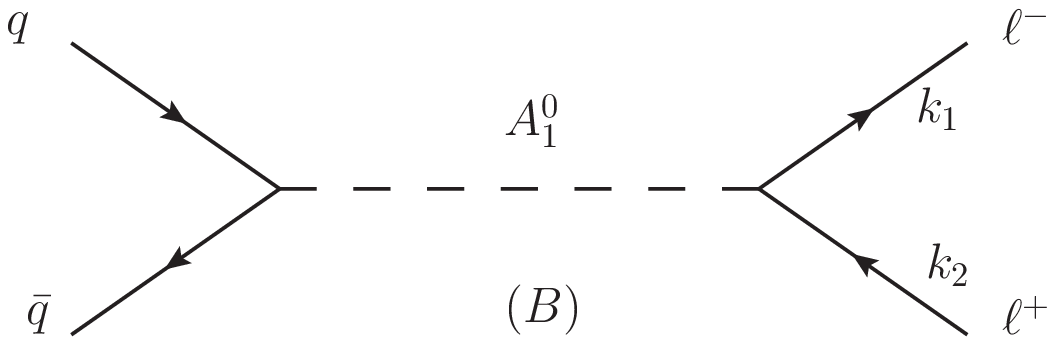}
\caption{Tree level contributions to the $\eta\to\ell^+\ell^-$
decay}\label{Z0-A10}
\end{figure}

Besides the QED contribution, there exists a tree level
contributions induced by exchanging weak interaction gauge boson
$Z^0$ and a new CP-odd pseudoscalar boson $A_1^0$ at s-channel and
the Feynman diagrams are shown in Fig.(\ref{Z0-A10}).
\begin{eqnarray}
\calM_{Z^0}=\langle\ell^+\ell^-|\bar{\ell}V_{\bar{\ell}\ell
Z^0}\frac{-i}{p^2-m_Z^2+i\varepsilon}
\bar{q}V_{\bar{q}qZ^0}q|\eta\rangle\;,
\end{eqnarray}
where $q$ stands for the light quarks $u$, $d$ and $s$, $p=k_1+k_2$,
$V_{\bar{\ell}\ell Z^0}$ and $V_{\bar{q}qZ^0}$ are the interaction
vertices of $\bar{\ell}\ell Z^0$ and
$\bar{q}qZ^0$\cite{Yang:2009kq}.

It is well known that the physical pseudoscalar particles $\eta$ and
$\eta^\prime$ are mixtures of the flavor eigenstates $\eta_q$ and
$\eta_s$\cite{Feldmann:1998vh}:
\begin{eqnarray}
\eta_q&=&\frac{1}{\sqrt{2}}(u\bar{u}+d\bar{d})\;,\\
\eta_s&=&s\bar{s}\;,
\end{eqnarray}
as
\begin{eqnarray}
\left(\begin{array}{c}\eta\\
\eta^\prime\end{array}\right)=\left(\begin{array}{cc}\cos\phi
&-\sin\phi\\ \sin\phi &
\cos\phi\end{array}\right)\left(\begin{array}{c}\eta_q \\
\eta_s\end{array}\right)\;,
\end{eqnarray}
where $\phi$ is the mixing angle.

Our conventions of the decay constants $f_{\eta}^q$ and $f_{\eta}^s$
are taken from Ref.\cite{Feldmann:1998vh,Pham:2007nt}.
\begin{eqnarray}
\langle0|J_{\mu5}^j|\eta\rangle=if_{\eta}^{j}p_\mu\;\;\;(j=q,s),\label{decayconstant}
\end{eqnarray}
where $q$ stands as the lighter quarks $u$ and $d$,
$f_{\eta}^q=\cos\phi f^q$, $f_{\eta}^s=-\sin\phi f^s$ and $p_\mu$ is
the four-momentum of $\eta$.

After a straightforward calculation, we obtain the contributions of
the weak interaction sector to the amplitude:
\begin{eqnarray}
\calM_{Z^0}^u&=&\frac{2\sqrt{2}e^2f_{\eta}^qm_{\ell}m_{\eta}}{(\sin\theta_W\cos\theta_W)^2}\frac{1}{p^2-m_Z^2}\;,\nonumber\\
\calM_{Z^0}^{d}&=&-\frac{2\sqrt{2}e^2f_{\eta}^qm_{\ell}m_{\eta}}{(\sin\theta_W\cos\theta_W)^2}\frac{1}{p^2-m_Z^2}\;,\label{MZ0}
\end{eqnarray}
where the projection operator for outgoing lepton pair is employed.

In the NMSSM, the light CP-odd pseudoscalar Higgs $A_1^0$ couples to
up-, down-type quarks and leptons. Following the general notation of
Ref.\cite{He:2006fr}, one can write the amplitude in terms of the
effective couplings Eqs.(\ref{NMSSM-Lagrangian-1}) and
(\ref{NMSSM-Lagrangian-2}). As generally suggested in literature
that by fitting available data, $\tan\beta$ takes a larger value,
thus the coupling constant $l_u$ in Eq.(\ref{NMSSM-Lagrangian-1}) is
much suppressed and  the contribution of $u-$type quarks to the
amplitude in the NP part is negligible. Then the extra contribution
of for $d-$type quarks to the amplitude  reads:
\begin{eqnarray}
\calM_{A_1^0}^d=-\frac{l_dg_\ell
f_{\eta}^qm_{\ell}m_{\eta}^2}{2\sqrt{2}v^2}\frac{1}{m_{\eta}^2-m_{A_1^0}^2+im_{A_1^0}\Gamma_{A_1^0}}\bar{u}(k_1)\gamma_5v(k_2)\;,\label{MA10}
\end{eqnarray}
where $\Gamma_{A_1^0}$ is the total width of $A_1^0$. When deriving
Eq.(\ref{MA10}), we utilize the relation,
\begin{eqnarray}
\langle0|\bar{d}\gamma_5
d|\eta\rangle=-i\frac{f_{\eta}^qm_\eta^2}{2\sqrt{2}m_d}\;,\;\;\langle0|\bar{s}\gamma_5
s|\eta\rangle=-i\frac{f_{\eta}^sm_\eta^2}{2m_s}\;,\label{decayconstant-1}
\end{eqnarray}
where $f_{\eta}^q$ and $f_{\eta}^s$ are defined in
Eq.(\ref{decayconstant}).

We can further reduce $\calM_{A_1^0}^d$ into:
\begin{eqnarray}
\calM_{A_1^0}^d=-\frac{l_dg_\ell
f_{\eta}^qm_{\ell}m_{\eta}^3}{2v^2}\frac{1}{m_{\eta}^2-m_{A_1^0}^2+im_{A_1^0}\Gamma_{A_1^0}}\;.\label{MA10-1}
\end{eqnarray}

By the aforementioned notation, one can easily obtain
$\calM_{A_1^0}^s$ as $\sqrt{2}\calM_{A_1^0}^d {f_{\eta}^s\over
f_{\eta}^q} $.

The total contribution is a sum of all the individual ones:
\begin{eqnarray}
\calM_{tot}=\calM_{\gamma\gamma}+\calM_{Z_0}^{u,d,s}+e^{i\theta_{NP}}\calM_{A_1^0}^{u,d,s}\;,
\end{eqnarray}
where $\theta_{NP}$ represents a possible relative phase between the
contributions of SM and NMSSM.

The total decay rate of $\eta\to\mu^+\mu^-$ is expressed as:
\begin{eqnarray}
\Gamma_{tot}(\eta\to\mu^+\mu^-)=\frac{1}{8\pi}\frac{|{\bf k}|}{m_\eta^2}|\calM_{tot}|^2\;,
\end{eqnarray}
where ${\bf k}$ is the three-momentum of one of the leptons in the
rest frame of $\eta$.
\subsection{For decays of heavy pseudoscalar mesons}
Generally, when calculating the anomaly and decay rate of $\pi^0\to
\gamma\gamma$, for simplification, one can use an approximation
$q\rightarrow 0$ \cite{Cheng:1985bj}. This approximation works well
for decays of light pseudoscalar mesons, but definitely not for
heavy pseudoscalar mesons with $q^2=M^2\gg 0$ where $M$ stands as
the mass of the heavy meson. Thus, we take another approach to take
into account the effects induced by the hadronic structure of the
decaying heavy meson.

\begin{figure}[h]
\centering
\includegraphics[width=12cm]{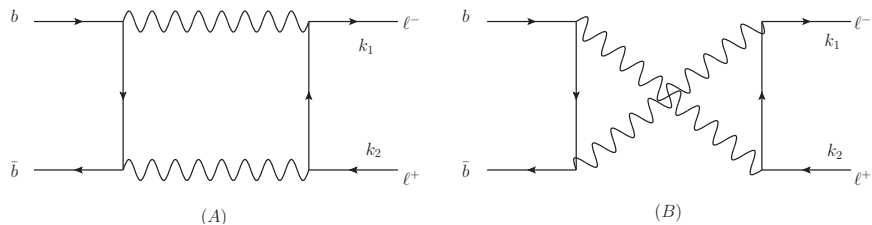}
\caption{The QED contributions to the $\eta_b\to\ell^+\ell^-$ decay
via box diagrams.}\label{oneloop-box}
\end{figure}

For decay of $\eta_b$ into lepton pairs, we employ the light-cone
distribution amplitude (LCDA) method to calculate the transition
amplitude  $\calM_{\gamma\gamma}$. The leading-order contributions
induced by the photon-Fermion loop are displayed in
Fig.(\ref{oneloop-box}).

Again, for completeness,  we re-derive the QED contribution to
$\eta_b\to \tau^+\tau^-$, and the corresponding formula was obtained
for $\eta_c\to \mu^+\mu^-$ in Ref.\cite{Yang:2009kq}. The
contribution of the loop diagrams is:
\begin{eqnarray}
\calM_{\gamma\gamma}&=&\calM_{\gamma\gamma}^A+\calM_{\gamma\gamma}^B\nonumber\\
&=&C_1\bar{u}(k_1)\gamma_5v(k_2)+C_2\bar{u}(k_1)\sigma^{\mu\nu}v(k_2)\varepsilon_{\mu\nu\rho\sigma}
k_1^\rho k_2^\sigma\;,\label{C1C2}
\end{eqnarray}
where
\begin{eqnarray}
C_1&=&\frac{Q_b^2e^4}{8\pi^2}f_{\eta_b}
m_{\ell}\int_0^1du\phi(u,\mu)\int_0^1dx\int_0^{1-x}dy\nonumber\\
&\times&\left\{
\frac{2}{D_1}+\frac{1+(2u-1)x-y}{D_2}+\frac{1+(1-2u)x-y}{D_3}\right\}\;,\\
C_2&=&\frac{Q_b^2e^4}{8\pi^2}f_{\eta_b}m_\ell\int_0^1du\phi(u,\mu)\int_0^1dx
\int_0^{1-x}\int_0^{1-x-y}dz\frac{1-x-y-z}{D_4^2}\;,
\end{eqnarray}

Here, the notions $D_{1,2,3,4}$ and the concrete expressions for
$C_1$ and $C_2$ are explicitly presented in Ref.\cite{Yang:2009kq}.
For $\eta_b\to\tau^+\tau^-$, the numerical results of $C_1$ and
$C_2$ are displayed in Table \ref{results-C1C2-cb}.

\begin{table}
\begin{center}
\renewcommand\arraystretch{1.4}
\begin{tabular}{|c|c|c|}
\hline\hline  &$C_1$ &$C_2$  \\
\hline $\eta_b\to \tau^+\tau^-$  &$4.22\times10^{-7}+5.44\times10^{-6}i$ &$-2.39\times10^{-8}+2.0\times10^{-8}i$ \\
\hline \hline
\end{tabular}
\end{center}\caption{The coefficients $C_1$ and $C_2$
in Eq.(\ref{C1C2})for
$\eta_{b}\to\tau^+\tau^-$.}\label{results-C1C2-cb}
\end{table}

Our next step would be evaluating the hadronic matrix element
$\langle0| \bar{b}_\alpha(x)b_\beta(y)|\eta_b\rangle$. Thus we need
the wave function for the pseudo-scalars. According to
Refs.\cite{Yang:2009kq,Chernyak:1983ej,Braun:1989iv}, we employ the
light-cone distribution amplitude for the calculation:
\begin{eqnarray}
\langle0|\bar{b}(x)_\alpha
b(y)_\beta|\eta_b(p)\rangle=-\frac{i}{4}f_{\eta_b}\int_0^1du e^{-i(u
p\cdot x+\bar{u}p\cdot
y)}\left[\pslash\gamma_5\right]_{\beta\alpha}\phi(u,\mu)\;,
\end{eqnarray}
where $\bar{u}=1-u$, $\mu$ is the energy scale and $f_{\eta_b}$ is
the decay constant of $\eta_b$ defined in Eq.(\ref{decayconstant}).
The wave function of $\eta_b$ is adopted as:
\begin{eqnarray}
\phi(u)=N4u(1-u)e^{-\frac{\beta}{4u(1-u)}}\;,
\end{eqnarray}
where $N$ is the normalization factor and the parameter is set as
$\beta=3.8\pm 0.7$ \cite{Yang:2009kq}.

\section{Numerical Analysis}
Firstly, we list some necessary input parameters which are taken
from either the PDG book\cite{Nakamura:2010zzi} or concerned
literatures \cite{Hwang:1997ie,:2009pz,Ke:2010tk,Rashed:2010jp}:
\begin{eqnarray}
\begin{aligned}
&f^q=1.07f_\pi,&& f^s=1.34 f_\pi,&&f_\pi=130\text{MeV},\nonumber\\
&f_{\eta_b}=0.705\text{MeV},&&m_{\eta_b}=9390.9\pm2.8\text{MeV},&&\Gamma_{\eta_b}=10\text{MeV},\nonumber\\
&\phi=(39.9\pm2.6(exp)\pm2.3(the))^{\circ}, &&\sin^2\theta_W=0.23116.&&
\end{aligned}
\end{eqnarray}

The reduced amplitude $\calA^{\ell}(m_{\eta}^2)$ in
Eq.(\ref{reduced-amplitude}) is:
\begin{eqnarray}
\calA^{\ell}(m_{\eta}^2)=2.61-5.21
i\;,\calA^{\ell}(m_{\eta^\prime}^2)=7.59+3.41 i\;.
\end{eqnarray}

For the New Physics part, we firstly discuss the relevant input
parameters. The electroweak scale $v$ is 246 GeV. Furthermore, the
coupling constant $|g_\ell|$ in Eq.(\ref{NMSSM-Lagrangian-2}) is
stringently constrained by the muon anomalous magnetic
moment\cite{He:2006fr}:
\begin{eqnarray}
|{g_\ell}|\lesssim 1.2.
\end{eqnarray}
This bound results in an $A_1^0$ width
$\Gamma_{A_1^0}\lesssim3.7\times10^{-7}$ MeV. That implies that the
coupling constant $l_d$ is also of order of unity which is
consistent with the general estimates for the size of
$v\delta_{-}/(\sqrt{2}x)$\cite{Hiller:2004ii}.
\begin{figure}[h]
\centering
\includegraphics[width=7cm]{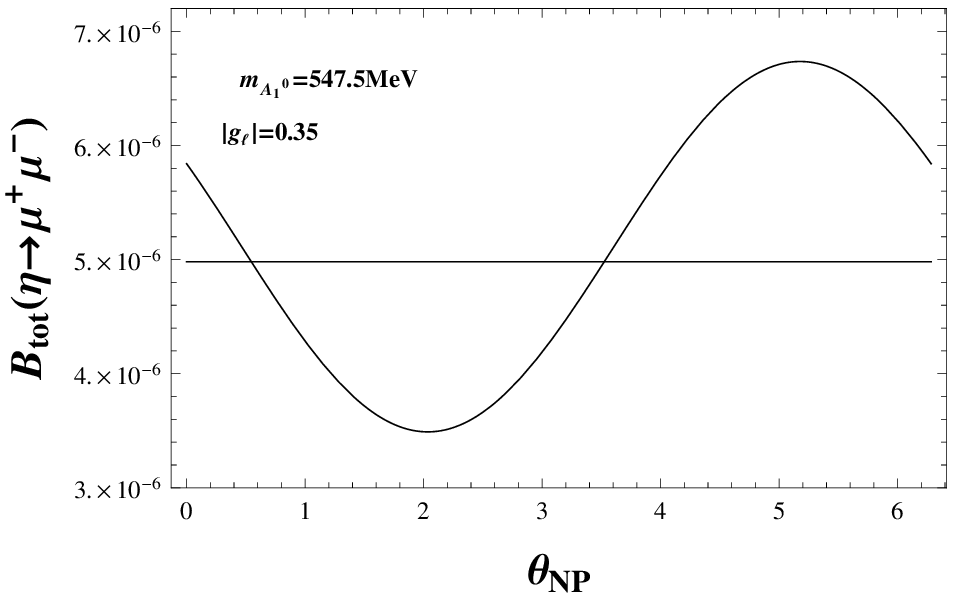}
\includegraphics[width=7cm]{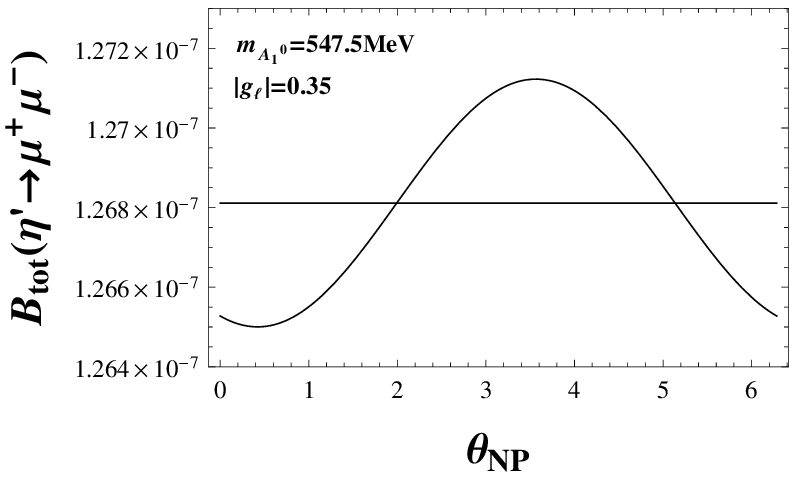}\\
\caption{The dependance of $\calB_{tot}(\eta\rightarrow\mu^+\mu^-)$
and  $\calB_{tot}(\eta^\prime\rightarrow\mu^+\mu^-)$ on the NP
phase. The horizontal sold lines correspond to the results of the
SM, respectively.}\label{eta1}
\end{figure}

Generally, a relative phase between the SM and NP pieces can exist
in the Lagrangian, and  $\theta_{NP}$ should be be determined by
fitting data or from a larger symmetry which includes the SM and the
concerned NP, in this work we treat it as a free parameter. We
illustrate the dependence of
$\calB_{tot}(\eta(\eta^\prime)\rightarrow \mu^+\mu^-)$ on this phase
in Fig.(\ref{eta1}). From this figure one can observe that, as
setting $|g_{\ell}|=0.35$, when $\theta_{NP}$ is around 0,
$\calB_{SM}(\eta\rightarrow\mu^+\mu^-)$ is enhanced by the NP
effect, whereas if $\theta_{NP}$ is about $\pi$, it decreases.
Comparing the result of $\eta\to\mu^+\mu^-$ with the experimental
data
\begin{eqnarray}
\calB_{Exp}(\eta\to
\mu^+\mu^-)&=&(5.8\pm0.8)\times10^{-6}\;.\label{experiment-data}
\end{eqnarray}
whose central value is above the SM prediction by about 20\%, we are
tempted to conclude that as the coupling constant is not very large
($|g_{\ell}|\lesssim 1.2$) as suggested in the literature, if one
expects to substantially enhance theoretical prediction of
$\calB_{SM}(\eta\rightarrow\mu^+\mu^-)$ to the experimental value
\cite{Nakamura:2010zzi} via the effect induced by $A_1^0$, the phase
$\theta_{NP}$ should be around $0$. Thus in following calculations
we set $\theta_{NP}$ to be $0$. By contrast,
$\calB_{SM}(\eta'\rightarrow\mu^+\mu^-)$ decreases near 0 and almost
reaches the maximum at $\pi$. Unfortunately, up to now, there is no
any experimental data on $\eta^\prime\to\mu^+\mu^-$,  with
$\theta_{NP}\approx 0$, its branching ratio is predicted, so that
the future experiments may hint us if the scenario works.

Therefore, by fitting the experimental data of
$\calB_{Exp}(\eta\rightarrow\mu^+\mu^-)=(5.8\pm0.8)\times10^{-6}$
while taking $|g_{\ell}|=0.35$ and $\theta_{NP}=0$, the proper mass
of $A_1^0$ should be $547.5$ MeV.

\begin{figure}[h]
\centering
\includegraphics[width=7cm]{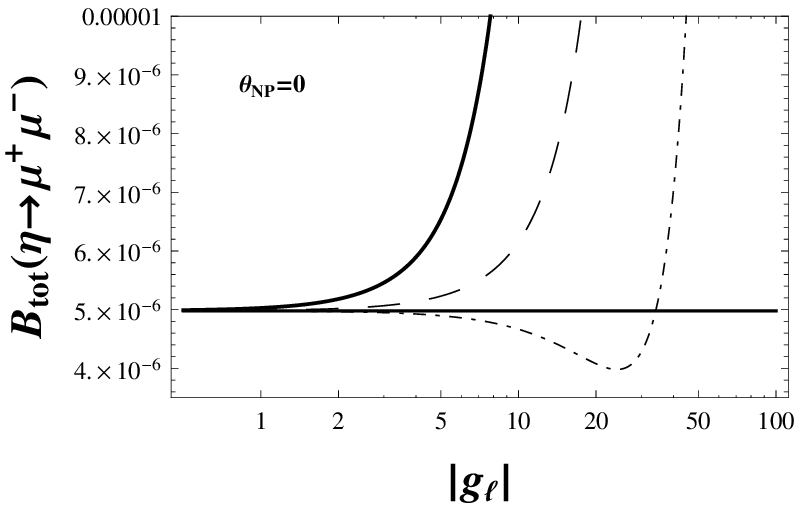}
\includegraphics[width=7cm]{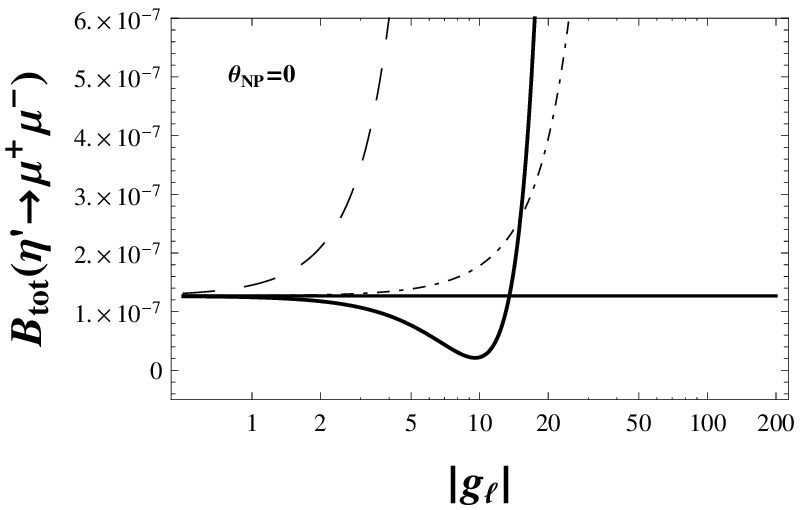}\\
\caption{The dependance of $\calB_{tot}(\eta\rightarrow\mu^+\mu^-)$
and $\calB_{tot}(\eta^\prime\rightarrow\mu^+\mu^-)$ on the coupling
constant $|g_{\ell}|$. For the $\eta$ decay, the dashed line, black
line and dot-dashed line correspond to $m_{A_1^0}=214.3$ MeV, 500
MeV and 1 GeV respectively. For the $\eta^\prime$ decay, the black
line, dashed line and dot-dashed line correspond to
$m_{A_1^0}=547.5$ MeV, 1 GeV, and 2 GeV respectively. The phase
angle is chosen as $\theta_{NP}=0$. The abscissas are the results of
the SM.} \label{eta3}
\end{figure}

In order to give a better insight, we draw the dependence of
$\calB_{tot}(\eta(\eta^\prime)\rightarrow\mu^+\mu^-)$ on
$|g_{\ell}|$ and $m_{A_1^0}$ in Fig.(\ref{eta3}).

From Fig.(\ref{eta3}), we observe that as $|g_{\ell}|$ is not very
large and $\theta_{NP}=0$, the NP effect for $\eta\to\mu^+\mu^-$
induced by existence of $A_1^0$ can enhance the branching ratio to
the experimental level, but a heavier $A_1^0$ with a mass about 1
GeV or more would not.

Based on these parameters employed in above text, we present all the
numerical results and corresponding experimental data for
$\eta(\eta^\prime)\to\mu^+\mu^-$ in Table.(\ref{eta-BR-SM-TOT}).

$\calB(\eta_c \to \mu^+\mu^-)$ in SM was studied in
Ref.\cite{Yang:2009kq} and its result is:
\begin{eqnarray}
\calB_{SM}(\eta_c \to
\mu^+\mu^-)=6.39^{+1.03}_{-0.89}\times10^{-9}\;.
\end{eqnarray}
With the same method we calculate $\calB(\eta_b\to\tau^+\tau^-)$ in
this work. However, since $\l_u=l_d/\tan^2\beta$,  for a larger
$\tan\beta$ which is usually considered in literature, the effect
induced by $A_1^0$ on $\calB(\eta_c\to \mu^+\mu^-)$ is negligible.
Indeed, if there were the experimental data for the decay of
$\eta_c\to\mu^+\mu^-$, one could gain more information about $A_1^0$
by comparing it with $\calB(\eta_b\to\tau^+\tau^-(\mu^+\mu^-))$.

\begin{figure}[h]
\centering
\includegraphics[width=7cm]{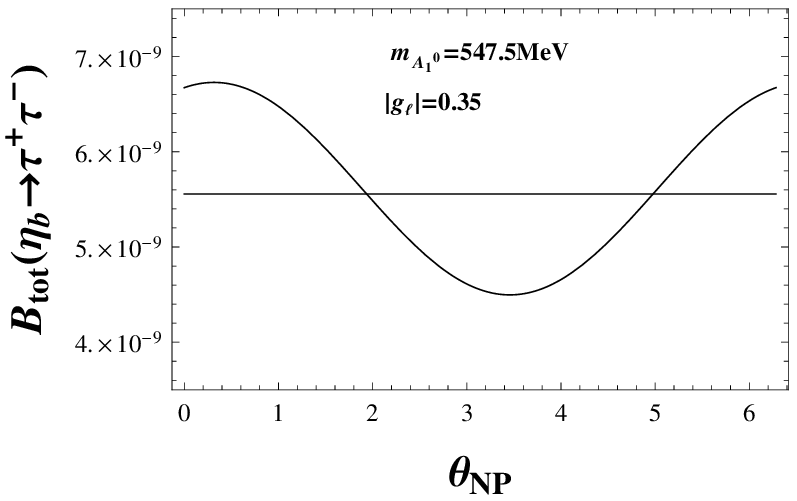}
\includegraphics[width=7cm]{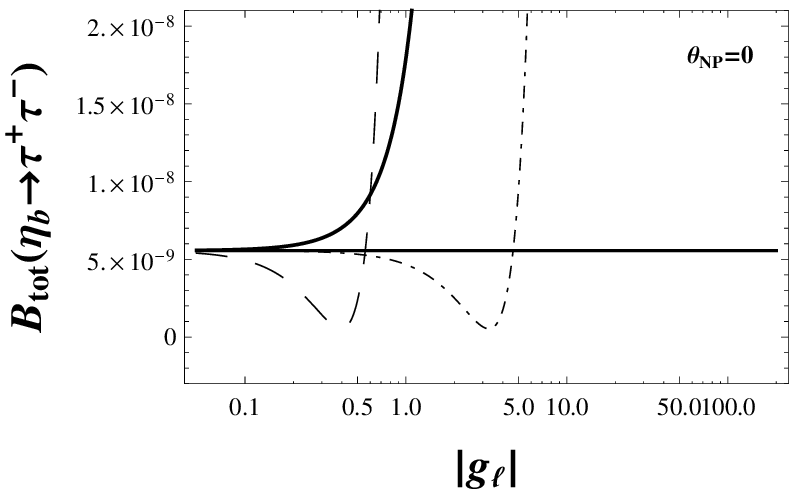}\\
\caption{(Left)The dependance of
$\calB_{tot}(\eta_b\rightarrow\tau^+\tau^-)$ on the NP phase angle.
The horizontal lines represent the results of the SM. (Right)The
dependance of $\calB_{tot}(\eta_b\rightarrow\tau^+\tau^-)$ on the
coupling cosntant$|g_{\ell}|$, where the black line, dashed line and
dot-dashed line represent the figures with $m_{A_1^0}=547.5$MeV,
$m_{A_1^0}=10$GeV£¬ and $m_{A_1^0}=$30GeV, respectively. The phase
angel is chosen as $\theta_{NP}=0$. The horizontal lines are the
results of the SM.}\label{etab13}
\end{figure}

Then, using $m_{A_1^0}=547.5$ MeV, we  obtain the branching ratio
predicted by the pure SM and NMSSM with this light CP-odd Higges
respectively. In analog to Fig.(\ref{eta1}) and Fig.(\ref{eta3}), we
draw Fig.(\ref{etab13}) for the heavy pseudoscalar meson $\eta_b$.
Additionally, we give the numerical results in Table
\ref{eta-BR-SM-TOT}.

Combining with the upshots of Ref.\cite{Chang:2008np} on  $\pi^0\to
e^+e^-$, we list all the results of the leptonic decays of the
pseudoscalar mesons $\pi^0,\;\eta,\;\eta^\prime,\;\eta_c,\;\eta_b$
in Table.(\ref{eta-BR-SM-TOT}).
\begin{table}[h]
\begin{center}
\renewcommand\arraystretch{1.4}
\begin{tabular}{|c|c|c|c|c|c|}
\hline\hline  &$\calB_{Exp}$ & $\calB_{SM}$
& $\calB_{tot}^{\theta=0}$ & $|g_{\ell}|$ & $m_{A_1^0}$(\small{MeV}) \\
\hline $\pi^0\to e^+e^-$  & $(7.48\pm0.38)\times10^{-8}$ & $(6.25\pm0.09)\times10^{-8}$ &$7.48\times10^{-8}$ &0.35 &134.95\\
\hline $\eta\to \mu^+\mu^-$  &$(5.8\pm0.8)\times10^{-6}$ & $4.94^{+0.44}_{-0.45}\times10^{-6}$ &$5.80^{fit}\times10^{-6}$ &0.35 &547.5\\
\hline $\eta^\prime\to \mu^+\mu^-$  & $-$ & $(1.27\pm0.42)\times10^{-7}$ &$(1.26\pm0.42)\times10^{-7}$ &0.35 &547.5\\
\hline $\eta_c\to \mu^+\mu^-$  &$-$ & $6.39^{+1.03}_{-0.89}\times10^{-9}$ &$6.39^{+1.03}_{-0.89}\times10^{-9}$ &0.35 &547.5\\
\hline $\eta_b\to \tau^+\tau^-$  &$<8\%$ & $5.56^{+0.44}_{-0.45}\times10^{-9}$ &$6.67^{+0.44}_{-0.45}\times10^{-9}$ &0.35 &547.5\\
\hline \hline
\end{tabular}
\end{center}\caption{The branching ratios of pseudoscalars to a leptonic pair in SM and in NMSSM.}\label{eta-BR-SM-TOT}
\end{table}

\begin{figure}[h]
\centering
\includegraphics[width=16cm]{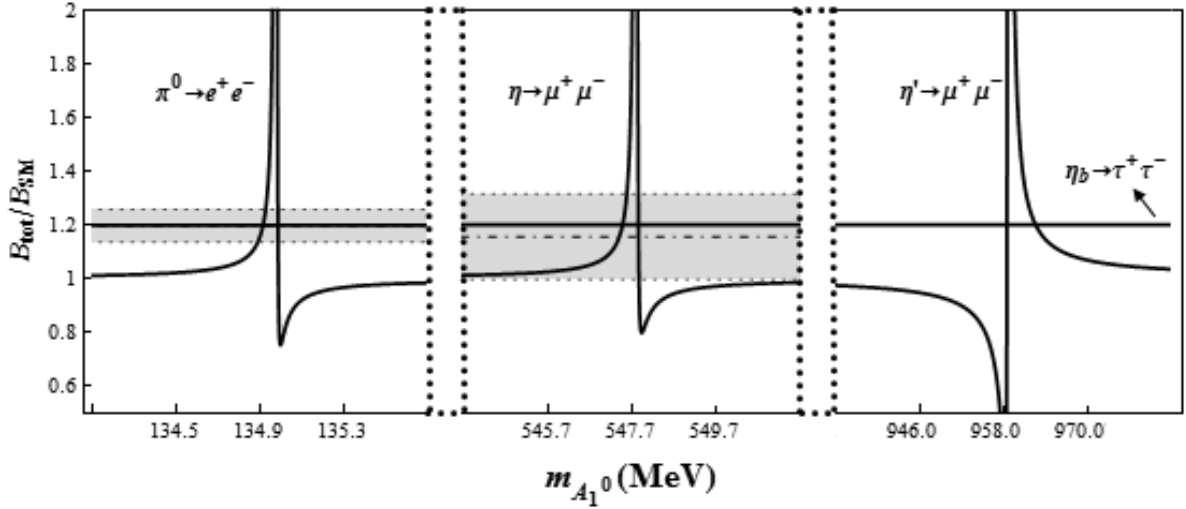}
\caption{Comparison among the leptonic decays of the pseudoscalar
mesons, where we have normalized each branching ratio with its SM
prediction. The dot-dash line represents the experimental data and
the shadowed region corresponds to the error band, where
$|g_\ell|=0.35$.}\label{compare}
\end{figure}

\section{Discussion and Conclusion}

The starting point of this work is that the SM prediction on
$\calB(\pi^0\to e^+e^-)$ is smaller than the data by a few standard
deviations, and the large rate of $\Sigma\to p+\mu^+\mu^-$ is also
beyond the SM prediction, thus there must be something new. But what
is it, new physics or a mechanism hidden in the SM but not being
taken into account? Definitely, it is not due to the final state
interaction because the leptonic and semi-leptonic decays are not
contaminated by non-perturbative QCD effects. Thus people are
inclined to attribute the discrepancy to new physics effects. But
then what new physics could it be?

To explain the anomalously large decay rate of $\Sigma^+\to
p+\mu^+\mu^-$, He, Tandean and Valencia proposed a new mechanism
where a light CP-odd scalar boson $m_{A_1^0}=214.3$ MeV
exists\cite{He:2006fr}. Later, some authors studied the transition
$\pi^0\to e^+e^-$ in terms of the same mechanism and their result
indicates that the suggested mass cannot fit the
data\cite{Chang:2008np}. This discrepancy might be caused by
experimental error of $\Sigma^+\to p+\mu^+\mu^-$ because there were
only a few events. Whether the mechanism is a reasonable one
motivates us to investigate the transitions $\pi^0\to e^+e^-;\;
\eta\to \mu^+\mu^-;\;\eta^\prime\to \mu^+\mu^-;\; \eta_c\to
\mu^+\mu^-;\;\eta_b\to\tau^+\tau^-$ within the same framework.

Looking at Fig.(\ref{compare}), one can notice several aspects.

1. Only the mass of $A_1^0$ is close to the mass of the decaying
pseudoscalar boson, the branching ratio of the leptonic decay can be
remarkably enhanced. That is due to the Breit-Wigner form of the
propagator:
$${1\over q^2-m_{A_1^0}^2+im_{A_1^0}\Gamma_{A_1^0}},$$
and $q^2=m_{\pi^0,\eta,\eta',\eta_b}$.

2. Unless the mass of the light $A_1^0$ is close to the mass of the
decaying pseudoscalar, its contribution to the leptonic decay is not
sensitive to the mass of $A_1^0$ at all. On the right panel of
Fig.(\ref{compare}), the abscissa corresponds to the decay of
$\eta_b\rightarrow \tau^+\tau^-$, which does not vary with respect
to the change of $m_{A_1^0}$, even though the contribution of
$A_1^0$ exists.

3. The experimental central value of $\calB(\eta\to \mu^+\mu^-)$ is
larger than the SM predictions, so that it implies that there could
be additional contributions from some mechanisms which were not
taken into account or from new physics beyond SM. On other aspect,
the error is large, i.e. within two standard deviations, the SM
prediction still coincides with the data. Thus if we take the
central values seriously, we should search for new sources of the
deviation. To further study, more accurate measurements are
necessary. Moreover, so far, there are no data on $\eta'\to
\mu^+\mu^-$ available, the reason is obvious that $\Gamma(\eta'\to
\mu^+\mu^-)$ and $\Gamma(\eta\to \mu^+\mu^-)$ have the same order of
magnitude, but $\Gamma_{tot}(\eta')\gg \Gamma_{tot}(\eta)$.

4. The existence of a light $A_1^0$ might be the source, but our
numerical results on the various leptonic decays of
$\eta,\eta',\eta_b$ as well as $\pi^0$ indicate that we cannot find
an universal mass for $A_1^0$ which can make the gaps between SM
predictions and data for $\Gamma(\pi^0\to e^+e^-)$ and
$\Gamma(\eta\to \mu^+\mu^-)$ simultaneously.

5. Just as Chang and Yang indicated, if the mass of $A_1^0$ is close
to pion mass, the gap between theoretical prediction and data may be
filled out, but another serious problem is raised. Namely, as
$\pi^0$ and $A_1^0$ have close masses, they should maximally mix
according to the general principle of quantum mechanics, if so, the
data on $\pi^0\to \gamma\gamma$ would not be explained. Similarly,
the argument can be applied to other pseudoscalar mesons. However,
from another aspect, as $\pi^0$ and $A_1^0$ have the same CP
behavior and are close in mass, in the two-photon final state, it is
hard to distinguish between them.

6. It is noted that the concerned experiments have larger errors, so
that one may re-consider if we can reconcile the experimental data
and  theoretical predictions with the help of $A_1^0$. Since the
measurement of $\pi^0\to e^+e^-$ has a smaller error, let us assume
that the mass of $A_1^0$ obtained by fitting $\calB(\pi^0\to
e^+e^-)$ is the right one, then with this value we re-examine
$\calB(\Sigma\to p+\mu^+\mu^-)$ and we obtain it as $1.16\sim
1.19\times 10^{-7}$. Using the measured data $(3.1^{+2.4}_{-1.9}\pm
1.5)\times 10^{-8}$, He, Tandean and Valencia got $m_{A_1^0}$ as
214.3 MeV. The estimated branching ratio with a lighter $A_1^0$ is 4
to 5 times larger than what they estimated. As pointed above, there
were only a few events, the errors may be large, so that we hope
that our experimentalists can strive to obtain more accurate
measurements on $\calB(\Sigma\to p+\mu^+\mu^-)$ which may provide
valuable information about $A_1^0$. Moreover, if $m_{A_1^0}$ is
close to 140 MeV, as we show in Fig.(\ref{compare}), its
contribution to the amplitude of $\eta\to\mu^+\mu^-$ is almost a
constant and negligible because the coupling of $A_1^0$ to
$\mu^+\mu^-$ is proportional to $m_{\mu}$ which is small compared to
$m_{\tau}$, meanwhile the measurement on $\eta\to \mu^+\mu^-$ also
possesses a larger error range and within 2 standard deviations the
SM prediction is consistent with the data, thus that data cannot
exclude an $A_1^0$ of about 140 MeV. There are so far, no data for
$\eta'\to \mu^+\mu^-,\,\eta_c\to \mu^+\mu^-$ and $\eta_b\to
\tau^+\tau^-$ available, as we show, an $A_1^0$ of about 140 MeV
does influence their branching ratios, but remains as a constant.
For $\eta_b\to \tau^+\tau^-$, the result of SM+NP is about 1.2 times
larger than the SM prediction. By contrast, as discussed in the
introduction, it does not affect $\eta_c\to \mu^+\mu^-$ at all.

Therefore, much more accurate measurements on $\Sigma\to
p+\mu^+\mu^-$ and the leptonic decays of the pseudoscalar mesons are
indeed badly needed.

As a conclusion, the phenomenology seems not to favor the light
CP-odd $A_1^0$, even though does not exclude its existence and there
exists a narrow window.

\vspace{0.7cm} {\bf Acknowledgments} \vspace{0.3cm}

We thank M.Z. Yang and Y.M. Wang for helpful discussion. The authors
would like to thank X.G. He and Y.D. Yang who explained their work
to us in some details. This work is supported by the National
Natural Science Foundation of China under contract No. 11075079 and
No. 11005079; the Special Grant for the Ph.D. program of Ministry of
Eduction of P.R. China No. 20100032120065.



\end{document}